\documentclass[12pt]{article}
\usepackage{epsfig, amssymb, graphicx, amsmath} 

\usepackage{enumitem}
\usepackage{eurosym}	
\usepackage{wrapfig}	
\usepackage{float}		
\usepackage{subfig}	
\usepackage{footnote}

\usepackage{bbm}		

\usepackage[authoryear]{natbib}
\bibpunct{(}{)}{,}{a}{}{,}

\newtheorem{proposition}{Proposition}

\newtheorem{lemma}{Lemma}

\newcommand{\un}{\underline}

\newcommand{\be}{\begin{equation}}
\newcommand{\en}{\end{equation}}
\newcommand{\ben}{\begin{equation*}}
\newcommand{\enn}{\end{equation*}}
\newcommand{\bea}{\begin{eqnarray}}
\newcommand{\ena}{\end{eqnarray}}

\begin{document}
 
\newlength\tindent
\setlength{\tindent}{\parindent}
\setlength{\parindent}{0pt}
\renewcommand{\indent}{\hspace*{\tindent}}

\begin{savenotes}
\title{
\bf{ Back-of-the-envelope swaptions \\ in a very parsimonious \\
multicurve interest rate model
}}
\author{
Roberto Baviera$^\dagger$ 
}

\maketitle

\vspace*{0.11truein}
\begin{tabular}{ll}
$(\dagger)$ &  Politecnico di Milano, Department of Mathematics, 32 p.zza L. da Vinci, Milano \\
\end{tabular}
\end{savenotes}

\vspace*{0.11truein}

\begin{abstract}
\noindent
We propose an elementary model to price European physical delivery swaptions in multicurve setting with a simple exact closed formula.
The proposed model is very parsimonious: it  is a three-parameter multicurve extension of the two-parameter \citet{HW1990} model.

\noindent
The model allows also to obtain simple formulas for all other plain vanilla Interest Rate derivatives. Calibration issues are discussed in detail.
\end{abstract}

\vspace*{0.11truein}
{\bf Keywords}: 
Multicurve interest rates, parsimonious modeling, calibration cascade.
\vspace*{0.11truein}

{\bf JEL Classification}: 
C51, 
G12. 

\vspace{6cm}
\begin{flushleft}
{\bf Address for correspondence:}\\
Roberto Baviera\\
Department of Mathematics \\
Politecnico di Milano\\
32 p.zza Leonardo da Vinci \\ 
I-20133 Milano, Italy \\
Tel. +39-02-2399 4630\\
Fax. +39-02-2399 4621\\
roberto.baviera@polimi.it
\end{flushleft}

\newpage

\begin{center}
\Large\bfseries 
 Back-of-the-envelope swaptions  \\ in a very parsimonious \\
multicurve interest rate model
\end{center}


\vspace*{0.21truein}


\section{Introduction}

The financial crisis of $2007$ has had a significant impact  also on Interest Rate (hereinafter IR) modeling perspective.
On the one hand, multicurve dynamics have been observed in main inter-bank markets (e.g. EUR and USD),
on the other volumes on exotic derivatives have considerably decreased and
 liquidity has significantly declined even on plain vanilla instruments.

\smallskip

While on the first issue there exist nowadays excellent textbooks \citep[see, e.g.][]{Henrard, Ruga}, 
the main consequence of the second issue, i.e. the need of very parsimonious models, has been largely forgotten in current financial literature
where the additional complexity of today financial markets is often faced with parameter-rich models.
In this paper the focus is on the two relevant issues of parsimony and calibration.

\bigskip

First, the parsimony feature is crucial: 
 in today (less liquid) markets one often needs to handle models with very few parameters both from a calibration and from a risk management perspective.

In this paper we focus on a three-parameter multicurve extension of the well known two-parameters \citet{HW1990} model. 
This choice is very parsimonious: 
one of the most parsimonious Multicurve HJM model in the existing literature is the one introduced by \citet{MoreniPallavicini} that, in the simplest WG2++ case, requires ten free parameters. Another one has been recently proposed 
by \citet{grbac}, that in the simplest model parametrization involves at least
seven parameters.

\smallskip

Second, the model should allow for a calibration cascade, the methodology followed by practitioners, that consists 
in calibrating first IR curves via bootstrap techniques and then volatility parameters. This cascade is crucial and the reason is related again to liquidity. 
Instruments used in bootstrap, as FRAs, Short-Term-Interest-Rate (STIR)  futures and swaps, are several order of magnitude more liquid than the corresponding options on these instruments.

The proposed model,
besides the calibration of the initial {\it discount} and {\it pseudo-discount} curves, allows to 
price with exact and simple closed formulas all plain vanilla IR options: caps/floors, STIR options and European swaptions.
While caps/floors and STIR options can be priced with straightforward modifications of solutions already present in the literature 
\citep[see, e.g.][]{HenrardCrisis, BavieraCassaro},
in this paper we focus on pricing European physical delivery swaption derivatives (hereinafter swaptions). 

We also show in a detailed example the calibration cascade, where the volatility parameters are calibrated via swaptions.

\bigskip

The remainder of the paper is organized as follows.
In Section 2, we recall 
the characteristics of a swaption derivative contract in a 
general multicurve setting.
In Section 3 we introduce the Multicurve HJM framework and the parsimonious model within this framework; we also prove model swaption closed formula.
In section 4 we show in detail model calibration.
Section 5 concludes.

\section{Interest Rate Swaptions in a multicurve setting}
\label{sec:Swaption}


Multicurve setting for interest rates can be found in the two textbooks of \citet{Henrard} and \citet{Ruga}.
In this section we briefly recall interest rate notation and some key relations, with a focus on swaption pricing in a multicurve setting.

Let $( \Omega, {\cal F}, {\mathbb{P}} )$, with $\{ {\cal F}_t : t_0 \le t \le T^*  \}$,
be a complete filtered probability space satisfying the usual hypothesis, where $t_0$ is the value date and $T^*$ a finite time horizon for
all market activities.
Let us define $B(t, T)$ the {\it discount} curve with $t_0 \le t<T < T^*$ and $D(t, T)$, the stochastic discount, s.t.
\be
B(t, T) = \mathbb{E} \left[ D(t, T) | {\cal F}_t \right] \;\; .
\label{eq:base}
\en
The quantity $B(t, T)$ is often called also risk-free zero-coupon bond.
For example, market standard in the Euro interbank market is to consider as {\it discount} curve the EONIA curve (also called OIS curve).
As in standard single curve models, forward discount $B(t; T, T + \Delta)$ is equal to the ratio $B(t, T + \Delta)/B(t, T)$. 
A consequence of (\ref{eq:base}) is that $B(t; T, T + \Delta)$ is a martingale in the $T$-forward measure.\footnote{The $T$-forward measure is defined as the probability measure s.t. 
$B(t, T) \, \mathbb{E}^{(T)} \left[ \, \bullet  \, | {\cal F}_t \right] = \mathbb{E} \left[ D(t, T) \, \bullet  | {\cal F}_t \right]$ 
\citep[see, e.g.][]{Musiela}. }

\bigskip

As in \citet{Henrard}, 
also a {\it pseudo-discount} curve is considered. 
The following relation holds for Libor rates $L(T, T + \Delta) $ and the corresponding forward rates $ L(t; T, T + \Delta)$ in $t$
\be
B(t, T + \Delta) \, L(t; T, T + \Delta) := \mathbb{E} \left[ D(t, T + \Delta) \, L(T, T + \Delta)  | {\cal F}_t \right] \; ,
\label{eq:ForwardRate}
\en
where the lag $\Delta$ is the one that characterizes the {\it pseudo-discount} curve; 
e.g. $6$-months in the Euribor $6$m case.

The (foward) {\it pseudo-discounts} are defined as
\be
\hat{B}(t; T, T + \Delta) \,  := \frac{1}{1 +  \delta(T, T + \Delta) \, L(t; T, T + \Delta)}  
\label{eq:pseudo}
\en
with $\delta(T, T + \Delta)$ the year-fraction between the two calculation dates for a Libor rate  
and the {\it spread} is defined as
\[
\beta(t; T, T + \Delta) \,  := \frac{B(t; T, T + \Delta)}{\hat{B}(t; T, T + \Delta)} \;\; .
\]

From equation (\ref{eq:ForwardRate}) one gets
\be
B(t, T) \, \beta(t; T, T + \Delta) = \mathbb{E} \left[ D(t, T) \, \beta(T, T + \Delta)  | {\cal F}_t \right]
\label{eq:condition}
\en
i.e. $\beta(t; T, T + \Delta)$ is a martingale in the $T$-forward 
measure.
This is the unique property that process $\beta(t; T, T + \Delta)$ has to satisfy.

Hereinafter, as market standard, all discounts and OIS derivatives refer to the {\it discount} curve, 
while forward forward Libor rates are always related to the corresponding {\it pseudo-discount} curve via (\ref{eq:pseudo}).

\subsection{Swaption}

A swaption is a contract on the right to enter, at option's expiry date  $t_\alpha$, in a payer/receiver swap with a strike rate $K$ established 
when the contract is written.

The underlying swap at expiry date  $t_\alpha$ is composed by a floating and a fixed leg; 
typically payments do not occur with the same frequency in the two legs (and they can have also different daycount) and 
this fact complicates the notation. Flows end at swap maturity date $t_{\omega}$.
We indicate floating leg payment dates as $\mathbf{t}' := \left\{ t'_\iota \right\}_{\iota=\alpha' +1 \ldots \omega'}$ 
(in the Euro market, typically versus Euribor-6m with semiannual frequency and Act/360 daycount),
and fixed leg payment dates $\mathbf{t} := \left\{ t_j \right\}_{j=\alpha +1 \ldots \omega}$ 
(in the Euro market, with annual frequency and 30/360 daycount);
we define also $t'_{\alpha'} := t_{\alpha}$, $t'_{\omega'} := t_{\omega}$.

Let us introduce the following shorthands
\[
\left\{
\begin{array}{lcl}
B_{\alpha \, j} (t) & := &  B (t; t_{\alpha}, t_{j}) \\[2mm]
B_{\alpha' \, \iota} (t) & := &  B (t; t'_{\alpha'}, t'_{\iota}) \\[2mm]
\beta_{\iota} (t) & := &  \beta (t; t'_\iota, t'_{\iota+1}) \\[2mm]
\delta'_\iota & := & \delta (t'_\iota, t'_{\iota+1}) \\[2mm]
\delta_j & := & \delta (t_j, t_{j+1}) \\[2mm]
c_j & := & \delta_j \; K \;\;\;  {\rm for} \;\;\; j = \alpha +1, \ldots, \omega -1 \;\;\; {\rm and} \;\;\; 1 +   \delta_\omega \; K \;\;\; {\rm for} \;\;\; j = \omega
\end{array}
\right. \; \; .
\]

A swap rate forward start  in $t_\alpha$ and valued in  $t \in [t_0, t_\alpha]$, 
$ S_{\alpha \omega}(t)$, 
is obtained equating in $t$ the Net-Present-Value 
of the floating leg and of the fixed leg
\[
 S_{\alpha \omega}(t) = \frac{ {\cal N}_{\alpha \omega}(t)}{ BPV_{\alpha \omega}(t)}
\]
with  the forward Basis Point Value 
\be
 {BPV}_{\alpha \omega}(t) := \sum^{\omega - 1}_{j = \alpha} 
  \delta_j \,  B_{\alpha \, j+1} (t)  \;
\en
and the numerator equal to the expected value in $t$ of swap's floating leg flows
\be
 {\cal N}_{\alpha \omega}(t) := \mathbb{E} \left[ \sum^{\omega' - 1}_{\iota = \alpha'} 
D (t, t'_ {\iota+1})  \, \delta'_\iota \, L(t'_\iota, t'_{\iota+1}) \bigg| {\cal F}_t \right] = 1 - B(t, t_\omega)  + 
\sum^{\omega' - 1}_{\iota = \alpha'} 
B (t, t'_\iota)  \left[   \beta_\iota(t) - 1 \right] \; ,
\en
where the last equality is obtained using relations  (\ref{eq:base}) and (\ref{eq:condition}).
Let us observe that the sum of floating leg flows is composed by two parts: the term $\left[ 1 - B(t, t_\omega) \right]$, equal to the single curve case, and
the remaining sum of $ B (t, t'_\iota) \left[  \beta_\iota(t) - 1 \right]$ that corresponds to the {\it spread} correction present in the multicurve setting.

Receiver swaption payoff at expiry date is
\be
{\cal{R}}_{\alpha \omega}(t_\alpha)  :=  BPV_{\alpha \omega}(t_\alpha)  
\left[  K - S_{\alpha \omega}(t_\alpha) \right]^+   = \left[  K \, BPV_{\alpha \omega}(t_\alpha)   - {\cal N}_{\alpha \omega}(t_\alpha) 
\right]^+ \; .
\label{eq:receiverPayoff}
\en

A receiver swaption is the expected value at value date of the discounted payoff
\[
\mathfrak{R}_{\alpha \omega}(t_0) := \mathbb{E} \left\{ D (t_0, t_\alpha) \, {\cal{R}}_{\alpha \omega}(t_\alpha) | {\cal F}_{t_0} \right\} =  B (t_0, t_\alpha) \, \mathbb{E}^{(\alpha)} \left\{ {\cal{R}}_{\alpha \omega}(t_\alpha) | {\cal F}_{t_0} \right\}
\]
where we have also rewritten the expectation in the $t_\alpha$-forward measure.


\begin{lemma} The two following two properties hold
\begin{enumerate}[label=\roman*)]
\item  $ {\cal N}_{\alpha \omega}(t) $ and $BPV_{\alpha \omega}(t) $ are martingale processes in the $t_\alpha$-forward measure for $t\in[t_0, t_\alpha]$;
\item Receiver swaption  payoff {\rm (\ref{eq:receiverPayoff})} reads
\be
\begin{split}
{\cal{R}}_{\alpha \omega}(t_\alpha) =
& \left[ B(t_\alpha, t_\omega) + K \, BPV_{\alpha \omega}(t_\alpha)  + 
\sum^{\omega' - 1}_{\iota = \alpha'} 
B (t_{\alpha}, t'_\iota)  \left[  1 - \beta_\iota(t_\alpha) \right]   - 1 
\right]^+ =  \\[1ex] 
& \left[ 
\sum^{\omega}_{j=\alpha + 1} c_j B_{\alpha j} (t_\alpha)  +
\sum^{\omega' -1}_{\iota=\alpha' +1} B_{\alpha' \iota} (t_\alpha)  - 
\sum^{\omega' -1}_{\iota=\alpha'} \beta_{\iota} (t_\alpha) \, B_{\alpha' \iota} (t_\alpha)  
\right]^+ 
\end{split}
\label{eq:payoff}
\en
\end{enumerate}
\label{le:payoff}
\end{lemma}
{\it Proof}.
Straightforward given the definitions of {\it discount} and {\it pseudo-discount} curves $\; \clubsuit$

\bigskip

This lemma has some relevant consequences.
On the one hand, property i) 
allows generalizing the Swap Market Model approach in \citep{jamshidian1997} to swaptions in the multicurve case, hence it allows obtaining market swaption formulas choosing properly the volatility structure. 
One can get the Black, Bachelier or Shifted-Black market formula \citep[see, e.g.][]{BM} where flows are discounted with the {\it discount} curve and
forward Libor rates are related to {\it pseudo-discounts} via (\ref{eq:pseudo}), as considered in market formulas. 
Moreover, property i) implies also that put-call parity holds also for swaptions in a multicurve setting. 

On the other hand, property ii) clarifies that a complete specification of the model for swaption pricing requires only the dynamics for
the forward {\it discount} and {\it spread} curves as specified in the next section.

\section{A Multicurve Gaussian HJM model with closed form swaption solution}

\bigskip

A Multicurve HJM model (hereinafter MHJM) 
is specified providing initial conditions for the {\it discount} curve $B(t_0, T)$ and the {\it spread} curve $\beta(t_0; T, T + \Delta)$, 
and indicating their dynamics.  
{\it Discount} and {\it spread} curves' dynamics in the MHJM framework we consider in this paper are
\be
\left\{
\begin{array}{lcll}
d  B (t; t_{\alpha}, t_{i}) & = & - B (t; t_{\alpha}, t_{i}) \left[ \sigma(t, t_i) - \sigma(t, t_\alpha) \right] \cdot \left[ d \un{W}_{\, t} + \rho \, \sigma(t, t_\alpha) \, dt \right] & 
t \in [t_0, t_\alpha] \\[2mm]
d  \beta(t; t_i, t_{i+1}) & = &   \beta(t; t_i, t_{i+1}) \left[ \eta(t, t_{i+1}) - \eta(t, t_{i}) \right] \cdot \left[ d \un{W}_{\, t} + \rho \, \sigma(t, t_i) \, dt \right] & 
t \in [t_0, t_i]
\end{array}
\right.
\label{eq:MHJM}
\en
where  $ \sigma(t,T) $ and $\eta(t, T)$ are d-dimensional
vectors of adapted processes (in particular in the Gaussian case they are deterministic functions of time) with 
$ \sigma(t,t) = \eta(t,t) = 0 $, $x \cdot y$ is the canonical scalar product between $x, y \in \Re^d$, 
 and $ \un{W} $ is a d-dimensional Brownian motion
with instantaneous covariance 
 $\rho = (\rho_{i \; j=1,..,d})$
\[
 d W_{i, t} \, dW_{j, t} = \rho_{i \, j} \, dt \;. 
\]

\bigskip

Model (\ref{eq:MHJM}) is the most natural extension of the single-curve  \citet{HJM}  model.
The first equation in (\ref{eq:MHJM}) corresponds to the usual HJM model for the {\it discount} curve \citep[see, e.g.][]{Musiela}. 
The second equation in (\ref{eq:MHJM})  is a very general continuous process satisfying condition (\ref{eq:condition}) for the {\it spread}. 

We do not impose any other additional condition for curves' dynamics as the independence hypothesis in \citet{Henrard} 
or the orthogonality condition in \citet{BavieraCassaro}.

\bigskip

Change of measures are standard in this framework, because they are a straightforward generalization of
single curve modeling approaches \citep[see, e.g.][]{Musiela}.
The process
\[
d \un{W}^{(i)}_{\, t}  :=  d \un{W}_{\, t} + \rho \, \sigma(t, t_i) \, dt \; \; 
\]
is a d-dimensional Brownian motion in the $t_i$-forward measure. It is immediate to prove that, given dynamics (\ref{eq:MHJM}),
$B (t; t_{\alpha}, t_{i})$ is martingale in the $ t_{\alpha}$-forward measure and
$ \beta(t; t_i, t_{i+1}) $ is martingale in the $ t_{i}$-forward measure.

\bigskip

{\it Remark 1}.
Given equations (\ref{eq:MHJM}), the dynamics for the {\it pseudo-discounts} (\ref{eq:pseudo}) in the $t_i$-forward measure is
\[
d \hat{B}(t; t_i, t_{i+1}) = - \hat{B} (t; t_i, t_{i+1}) \left[ \sigma_i(t) + \eta_i(t) \right] \cdot \left[ d \un{W}^{(i)}_{\, t} - \rho \, \eta_i(t) \, dt \right] \qquad t \in [t_0, t_i] 
\]
where $\sigma_i(t)  := \sigma(t, t_{i+1}) - \sigma(t, t_{i}) $ and $\eta_i(t) := \eta(t, t_{i+1}) - \eta(t, t_{i})$. 
The {\it pseudo-discount} 
has a volatility which is the sum of {\it discount} volatility $\sigma_i(t)$ and of {\it spread} volatility $\eta_i(t)$. 

\bigskip

In this paper 
we consider an elementary 1-dimensional Gaussian model within MHJM framework (\ref{eq:MHJM}).
Volatilities for the {\it discount} curve $\sigma(t,T)$ and for the {\it spread} curve $\eta(t,T)$  are modeled as
\be
\left\{
\begin{array}{lccl}
\sigma(t,T) & = & (1- \gamma) & v(t,T) \\[2mm]
\eta(t,T) &=& \gamma & v(t,T) 
\end{array}\right.
\qquad \displaystyle  {\rm with} \; \;  v(t,T) := 
\left\{
\begin{array}{ll} \displaystyle 
\sigma \, \frac{1 - e^{-a (T-t)}}{a} & a \in \Re^+\setminus\{0\} \\[2mm]
\sigma \,  (T-t)  & a = 0
\end{array}\right. \; 
\label{eq:model}
\en
with $a, \sigma \in \Re^+$ and $\gamma \in [0,1]$, the three model parameters. 

\bigskip

This model is the most parsimonious (non-trivial) extension of \citet{HW1990} to multicurve dynamics, for this reason 
we call it Multicurve Hull White (hereinafter MHW) model. 
For all parameters choices volatility $ v(t,T) $ is strictly positive.

\bigskip

The selection of this model originates from two facts related to  the IR derivatives available for calibration.
On the one hand, in the calibration cascade, ``linear" IR derivatives (i.e depos, FRAs, STIR futures and swaps) are used for 
{\it discount} and {\it pseudo-discount} initial curve bootstrap, while 
the other parameters are calibrated on IR options. In the market, liquid IR option are STIR options, caps/floors and swaptions; 
unfortunately options on OIS are not liquid in the market place \citep[see, e.g.][and references therein]{MoreniPallavicini}. 

On the other hand,  in liquid  IR options, the key driver is the pseudo-discount curve $\hat{B}(t,T)$ 
via a Libor rate or a swap rate, where the latter can be seen as combinations of Libor rates \citep[see, e.g. eq.(1.28) in][]{Ruga}. 
Hence, when IR curves move, the main driver is {\it pseudo-discount} curve, directly related to option underlyings; 
the {\it discount} curve appears only in weights or discount factors, and swaption sensitivities w.r.t. the {\it discount} curve are less than 
the corresponding sensitivities w.r.t. the {\it pseudo-discount} curve.

\bigskip

These facts lead to the conclusion that is much more difficult to calibrate volatility parameters specific to the {\it discount} curve. 
Thus
{\it Remark 1} plays a crucial role when  selecting the most parsimonious model within framework  (\ref{eq:MHJM}):
$\hat{B}(t,T)$ dynamics has volatility equal to $v(t,T)$ in MHW model  (\ref{eq:model}).
A parsimonious choice should associated a fraction $1- \gamma$ of volatility $v(t,T)$ to the {\it discount} curve and the remaining fraction $\gamma$ 
to the {\it spread} dynamics;
in fact, as previously discussed, options on OIS are not liquid enough  and then
a separate calibration of
$\sigma(t,T)$ and $\eta(t,T)$ in a generic MHJM is not feasible in practice.

\bigskip

Moreover, MHW model (\ref{eq:model}) allows pricing IR options in an elementary way. STIR options and caps/floors Black-like formulas can be obtained via a straightforward generalization of  the solutions in \citet{HenrardCrisis} and \citet{BavieraCassaro}. 
In this section we show that it is possible to price also swaptions via a simple closed formula. 
To the best of our knowledge, MHW model (\ref{eq:model}) is the first Multicurve HJM where all plain vanilla derivatives can be written with simple exact closed formulas
that are extensions of \citet{Black1976} formulas. 

\bigskip

The remaining part of this section is divided as follows.
We first show in {\bf Lemma \ref{le:xi}} how to write,  within MHW model (\ref{eq:model}), each element in receiver swaption payoff (\ref{eq:payoff})
as a simple function of one single Gaussian r.v. $\xi$. 
Then, (technical) {\bf Lemma \ref{le:unicity}} shows that swaption payoff can be rewritten as a function of $\xi$
and this function presents interesting properties. 
Finally in {\bf Proposition \ref{pr:swaption}}
 we prove the key result of this section: the exact closed formula for swaptions according to model (\ref{eq:model}).

\bigskip

It is useful to introduce the following shorthands
\[
\left\{
\begin{array}{lcll}
v_{\alpha' \, \iota} & := &  v(t_\alpha, t'_\iota)   & \iota = \alpha' , \ldots, \omega' \\[2mm]
\varsigma_{\alpha' \, \iota} & := &   (1- \gamma) \, v_{\alpha' \, \iota}   & \iota = \alpha' , \ldots, \omega \\[2mm]
\nu_{\alpha' \, \iota} & := & \varsigma_{\alpha' \, \iota} -   
\left( \eta(t_\alpha, t'_{\iota+1}) - \eta(t_\alpha, t'_{\iota})  \right) & \iota = \alpha' , \ldots, \omega'-1  \;\; .
\end{array}
\right.
\]

\bigskip

{\it Remark 2}. Volatilities $\{ v_{\alpha' \, \iota} \}_{\iota = \alpha' +  1 \ldots \omega'}$  
are always positive 
and  are  strictly  increasing with $\iota$.
The quantities $\{ \nu_{\alpha' \, \iota} \}_{\iota = \alpha' + 1 \ldots \omega'}$ can change sign depending on the value of $\gamma$.
In fact 
\[
\nu_{\alpha' \, \iota} =  v_{\alpha' \, \iota} - \gamma \, v_{\alpha' \, {\iota +1}} =  v_{\alpha' \, {\iota +1}}  \left( {\tilde \gamma}_\iota - \gamma \right)
\]
with ${\tilde \gamma}_\iota := v_{\alpha' \, \iota}/v_{\alpha' \, \iota+1} \in (0,1) $. 
Then, when $\gamma=0$ all $\{ \nu_{\alpha' \, \iota} \}_{\iota = \alpha' + 1 \ldots \omega' - 1}$ are positive and
$\nu_{\alpha' \, \alpha'} $ is negative, while for larger values of $\gamma$ some $\nu_{\alpha' \, \iota}$ become negative.
For $\gamma$ equal or close to $1$ all $\{ \nu_{\alpha' \, \iota} \}_{\iota = \alpha' \ldots \omega' - 1}$ are negative.
Due to these possible negative values, $\{ \nu_{\alpha' \, \iota} \}_{\iota}$ are not volatilities; we call them {\it extended} volatilities. 


\begin{lemma}
Discount and spread curves in $t_\alpha$ can be written, according to the MHW model {\rm (\ref{eq:model})} in the $t_\alpha$-forward measure, as
\be
\left\{
\begin{array}{rcrll}
B_{\alpha' \, \iota}(t_\alpha) & = &   B_{\alpha' \, \iota}(t_0) & 
\displaystyle \exp \left\{ -\varsigma_{\alpha' \, \iota}  \,  \xi  - \varsigma_{\alpha' \,\iota}^2 \, \frac{\zeta^2}{2} \right\} & \; \iota = \alpha' + 1, \ldots, \omega' \\[1ex]
\beta_\iota(t_\alpha) \, B_{\alpha' \, \iota}(t_\alpha) & = & \beta_\iota(t_0) \,  B_{\alpha' \, \iota}(t_0) &
\displaystyle \exp \left\{ -  \nu_{\alpha' \, \iota} \, \xi   - 
\nu_{\alpha' \, \iota}^2 \, \frac{\zeta^2}{2} \right\} & \; \iota = \alpha', \ldots, \omega' -1
\end{array}\right.
\label{eq:B dynamics}
\en
where 
\be
\xi :=  \int^{t_\alpha}_{t_0}  dW^{(\alpha)}_u \, e^{-a (t_\alpha-u)}
\label{eq:xi}
\en
a zero mean Gaussian r.v. whose variance is
\[
\zeta^2 :=  
\left\{
\begin{array}{ll}
\displaystyle \frac{1 - e^{-2 \, a (t_\alpha-t_0)}}{2 \, a}  & a \in \Re^+\setminus\{0 \} \\[2mm]
t_\alpha-t_0  & a  = 0 \; \; .
\end{array}\right.
\]
\label{le:xi}
\end{lemma}

{\it Proof}. 
A straightforward application of It\^{o} calculus, given dynamics (\ref{eq:MHJM}) and deterministic volatilities (\ref{eq:model}) 
$\; \clubsuit$

\bigskip 

A consequence of previous lemma is that
receiver swaption payoff (\ref{eq:payoff}) in the $t_\alpha$-forward measure
can be written as a function of a unique r.v. $\xi$ as
\be
{\cal{R}}_{\alpha \omega}(t_\alpha) =: \left[ f (\xi) \right]^+ \; .
\label{eq:Jam}
\en

In the following lemma we show that $f (\xi)$ is equal to a finite sum of exponential functions of $\xi$, i.e. 
\[
f (\xi)  = \sum_i w_i \, e^{\lambda_i  \xi} \; \; {\rm with} \; \; w_i, \lambda_i \in \Re
\]
where some $w_i < 0$ and some $\lambda_i \ge 0$. Hence, the swaption looks like a non-trivial spread option, with a number of terms equal to
$\omega - \alpha -1+ 2(\omega' - \alpha')$.

\bigskip

In {\bf Lemma \ref{le:unicity}} we prove that, even if the function $f$, for some parameters choices, is not a decreasing function of $\xi$, 
however
there exists a unique 
value $\xi^*$ s.t.  $f(\xi^*)=0$, 
i.e. the equality $ S_{\alpha \omega}(t_\alpha) = K$ is satisfied for this unique value.


\begin{lemma}
According to MHW model {\rm (\ref{eq:model})}, the function  $f (\xi)$ in swaption payoff is equal to 
\[
\begin{array}{rll}
f (\xi)  = 
& \displaystyle  \sum^{\omega}_{j=\alpha + 1} c_j B_{\alpha j} (t_0) \, e^{- \varsigma_{\alpha  j} \, \xi   -  \varsigma_{\alpha  j}^2 \, \zeta^2/2 } & {\rm (a)}\\[1ex]
& \displaystyle + \sum^{\omega' -1}_{\iota=\alpha' +1} B_{\alpha' \iota} (t_0) \, e^{- \varsigma_{\alpha'  \iota} \, \xi   -  \varsigma_{\alpha'  \iota}^2 \, \zeta^2/2}  & {\rm (b)} \\[1ex] 
& \displaystyle - \sum^{\omega' -1}_{\iota=\alpha'} \beta_{\iota} (t_0) \, B_{\alpha' \iota} (t_0) \, e^{- \nu_{\alpha'  \iota} \, \xi   -  \nu_{\alpha'  \iota}^2 \, \zeta^2/2}  & {\rm (c)} 
\end{array}
\]
and $\exists ! \, \xi^*$ s.t.  $f(\xi^*)=0$ for $a, \sigma \in \Re^+$ and $\gamma \in [0,1]$. 
Moreover, the function $f$ is greater than zero for $\xi < \xi^*$.
\label{le:unicity}
\end{lemma}

{\it Proof}.
See Appendix A $\; \clubsuit$

\bigskip


We have proven that,
even if function $f$ is not monotonic in its argument, there is a unique solution for equation $f(\xi) = 0$. This fact grants the possibility to
extend to MHW the approach of  \citet{Jamshidian}.
In the following proposition we prove that
a closed form solution holds for a receiver swaption for model (\ref{eq:model}).

\begin{proposition}
A receiver swaption, according to MHW model (\ref{eq:model}), can be computed with the closed formula 
\be
\begin{split}
\mathfrak{R}^\textsc{mhw}_{\alpha \omega}(t_0) = B(t_0,t_\alpha) 
& \Bigg\{ \displaystyle \sum^{\omega}_{j=\alpha + 1} c_j  \, B_{\alpha j} (t_0) \, N \left(\frac{\xi^*}{\zeta}+  \zeta \, \varsigma_{\alpha j}\right)   
+ \sum^{\omega' -1}_{\iota=\alpha' +1} B_{\alpha' \iota} (t_0)  \, N\left(\frac{\xi^*}{\zeta}+  \zeta \, \varsigma_{\alpha' \iota}\right)  \\
& \displaystyle - \sum^{\omega' -1}_{\iota=\alpha'} \, \beta_{\iota} (t_0) \, B_{\alpha' \iota} (t_0) \, N\left(\frac{\xi^*}{\zeta}+ \zeta \,  \nu_{\alpha' \iota}\right)  \Bigg\}  
\end{split}
\label{eq:MHWswaption}
\en
where $N(\bullet)$ is the standard normal CDF and $\xi^*$ is the unique solution of $f(\xi) =0$.
\label{pr:swaption}
\end{proposition}
{\it Proof}.
See Appendix A $\; \clubsuit$

\bigskip

Let us comment above proposition, which is the most relevant analytical result of this paper.
It generalizes the celebrated result of \citet{Jamshidian} to this Multicurve HJM model.
The main difference is that also negative addends appear in the receiver swaption $\mathfrak{R}^\textsc{mhw}_{\alpha \omega}(t_0)$
and there are {\it extended} volatilities instead of standard volatilities. 
It is straightforward to prove that, {\it mutatis mutandis}, a similar solution holds for a payer swaption.

\section{Model calibration}
\label{sec:Calibration}

In this section we show in detail model calibration of market parameters in the Euro market considering 
European ATM swaptions vs Euribor 6m
with the end-of-day market conditions of September 10, 2010 (value date). 

As discussed in the introduction, the calibration cascade is divided in two steps.
First, we bootstrap the {\it discount} and the {\it pseudo-discount} curves from 6m-Depo, three FRAs ($1 \times 7,  2 \times 8$ and $3 \times 9$)
and swaps (both OIS and vs Euribor 6m).
Then,  we calibrate the three MHW parameters ${\bf p} : =(a,\sigma,\gamma)$ 
with European ATM swaptions vs Euribor 6m on the 10y-diagonal 
(i.e. considering the $M=9$ ATM swaptions 1y9y, 2y8y, $\ldots$ , 9y1y). 

\begin{table}[h!]
\begin{center}
\begin{tabular}{|c||c|c|}
\hline
 & OIS rate (\%) & swap rate {\it vs} 6m (\%) \\
\hline
\hline
1w & -0.132 & - \\
\hline
2w & -0.132 & - \\
\hline
1m & -0.132 & - \\
\hline
2m & -0.133 & - \\
\hline
3m & -0.136 & - \\
\hline
6m & -0.139 & - \\
\hline
1y & -0.147 & 0.044 \\
\hline
2y & -0.135 & 0.080 \\
\hline
3y & -0.083 & 0.154 \\
\hline
4y & 0.008 & 0.259 \\
\hline
5y & 0.122 & 0.377 \\
\hline
6y & 0.254 & 0.512 \\
\hline
7y & 0.392 & 0.652 \\
\hline
8y & 0.529 & 0.786 \\
\hline
9y & 0.655 & 0.909 \\
\hline
10y & 0.766 & 1.016 \\
\hline
11y & 0.866 & 1.109 \\
\hline
12y & 0.957 & 1.195 \\
\hline
15y & 1.160 & 1.383 \\
\hline
\end{tabular}
\end{center}
\caption{\small OIS rates and  swap rates vs Euribor 6m in percentages: end-of-day mid quotes (annual 30/360 day-count  convention for  swaps vs 6m, 
Act/360 day-count for OIS) on 10 September 2015.}
\label{tab: Ois and Euribor data}	
\end{table} 

\begin{figure}
  	\begin{center}
      	\includegraphics[width=0.70\textwidth]{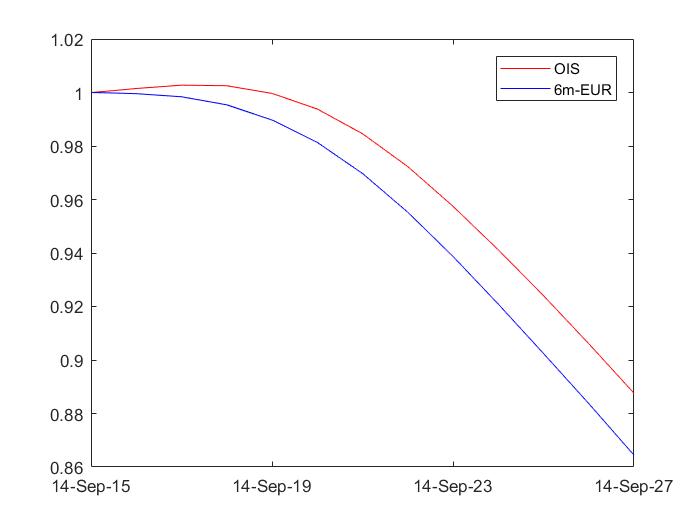}
	\end{center}
  	\vspace{-20pt}
  	\caption{\small {\it Discount} OIS curve (in red) and {\it pseudo-discount} Euribor-6m curve  (in blue)  on September 10, 2010, starting from the settlement date and up to a 12y time horizon.}
\label{fig1}
\end{figure}

The {\it discount} curve is bootstrapped from OIS quoted rates
with the same methodology described in \citet{BavieraCassaro}.
Their quotes at value date are reported 
in Table \ref{tab: Ois and Euribor data} (with market conventions, i.e.
annual payments and Act/360 day-count); in the same table we report also the swap rates
(annual fixed leg with 30/360 day-count). 
In Table \ref{tab:FRA and Euribor} we show the relevant FRA rates and the
Euribor 6m fixing  on the same value date (both with Act/360 day-count). All market data are provided by Bloomberg.  
Convexity adjustments for FRAs, present in the MHW model, are neglected because they do not impact the nodes relevant for the diagonal swaptions co-terminal 10y 
considered in this calibration and they are very small in any case.
In figure \ref{fig1} we show the {\it discount}  and {\it pseudo-discount} curves obtained via the bootstrapping technique.

\begin{table}[h!]
\begin{center}
\begin{tabular}{|c||c|}
\hline
 &  rate (\%) \\
\hline
\hline
Euribor 6m & 0.038 \\
\hline
FRA 1 $\times$ 7 & 0.038 \\
\hline
FRA 2 $\times$ 8 & 0.041 \\
\hline
FRA 3 $\times$ 9 & 0.043 \\
\hline
\end{tabular}
\end{center}
\caption{\small Euribor 6m fixing rate and FRA in percentages (day-count Act/360). FRA rates are end-of-day mid quotes at value date.}
\label{tab:FRA and Euribor}	
\end{table}

We show the swaption ATM volatilities in basis points (bps) in Table \ref{tab: Swaptions}; 
the swaption market prices are obtained according to the  standard normal market model; 
a model choice that allows for negative interest rates.

\begin{table}[h!]
\begin{center}
\begin{tabular}{|c|c|c|}
\hline
 expiry  & tenor & volatility (bps) \\
\hline
\hline
1y & 9y & 64.70 \\
\hline
2y & 8y &  66.78\\
\hline
3y & 7y &  68.53\\
\hline
4y & 6y &  70.91\\
\hline
5y & 5y &  72.36\\
\hline
6y & 4y &  73.07\\
\hline
7y & 3y &  73.21\\
\hline
8y & 2y &  73.51\\
\hline
9y & 1y &  73.45\\
\hline
\end{tabular}
\end{center}
\caption{\small  Normal volatilities for ATM diagonal swaptions co-terminal 10y in bps on 10 September 2015.}
\label{tab: Swaptions}	
\end{table} 

We minimize the square distance between swaption model and market prices 
\begin{equation*}
\mathtt{Err}^2({\bf p})=\sum_{i=1}^\textsc{M} \left[ \mathfrak{R}^\textsc{mhw}_i({\bf p}; t_0)-
\mathfrak{R}^\textsc{mkt}_i(t_0)  \right]^2
\end{equation*} 
where market ATM swaption pricing formula according to the multicurve normal model  is
reported in Appendix B.
 
\begin{figure}
  	\begin{center}
      	\includegraphics[width=0.70\textwidth]{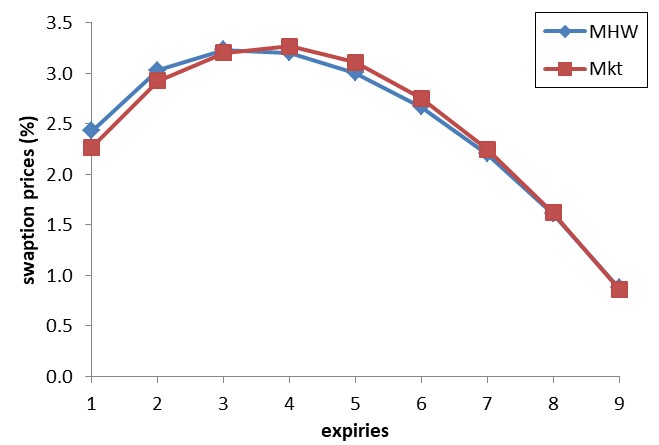}
	\end{center}
  	\vspace{-20pt}
  	\caption{\small Market prices for ATM diagonal swaptions co-terminal 10y in percentages (squares in red) and the corresponding ones obtained via the MHW calibration (diamonds in blue)
for the $9$ expiries considered.}
\label{fig2}
\end{figure}

We obtain the parameter estimations minimizing the $\mathtt{Err}$ function w.r.t. $a, \gamma$ and $\tilde \sigma := \sigma/a$; 
the solution is stable for a large class of starting points. 
As estimations we obtain $a=13.31 \%$, $\sigma=1.27 \%$ and $\gamma = 0.06 \%$.
The difference between model and market swaption prices are shown in figure \ref{fig2}:
calibration results look good despite the parsimony of the proposed model.

It is interesting to observe that the dependence of the $\mathtt{Err}$ function w.r.t. $\gamma$ is less pronounced compared to the one 
w.r.t. $a$ and $\sigma$; even if the minimum values for the $\mathtt{Err}$ function are 
achieved for very low values of $\gamma$, however, differences in terms of mean squared error are very small increasing,
even significantly, $\gamma$: 
another evidence 
that the most relevant dynamics for swaption valuation 
is the one related to the {\it pseudo-discount} curve, where the corresponding volatility 
 does not depend on $\gamma$ parameter. 

\section{Conclusions}

Is it possible to consider a parsimonious multicurve IR model without assuming constant {\it spreads}?
In this paper we introduce a three parameter generalization
of the two parameters \citet{HW1990} model, where the additional parameter $\gamma$  lies in the interval $[0,1]$.
The limiting cases correspond to some models already known in the literature: 
the case with $\gamma =0$ corresponds to the {\bf S0} hypothesis in \citet{HenrardCrisis}, where the {\it spread} curve is constant over time, 
while $\gamma =1$ corresponds to the {\bf S1} assumption in \citet{BavieraCassaro}.

\bigskip

We have proven that the model allows a very simple closed formula for European physical delivery swaptions (\ref{eq:MHWswaption})
with a formula, very similar to the one of \citet{Jamshidian}, 
with the presence of {\it extended} volatilities, that can assume negative values.
Model calibration is immediate: we have shown in detail 
how to implement the calibration cascade on the September 10, 2010 end-of-day market conditions. 

The proposed model allows also Black-like formulas for the other liquid IR options (caps/floors and STIR options) and
simple analytical convexity adjustments for FRAs and STIR futures; furthermore numerical techniques similar to the  
HW model can be applied.

\bigskip

This very parsimonious model is justified by 
the good calibration properties  on ATM swaption prices and 
by the observation that the {\it pseudo-discount} dynamics is the relevant one in the valuation of liquid IR options.  
Furthermore a very parsimonious model, as the proposed MHW model (\ref{eq:model}), can be the choice of election in 
challenging tasks where the multicurve IR dynamics is just one of the modeling elements: two significant examples are 
the pricing and the risk management of illiquid corporate bonds, 
and the XVA valuations including all contracts between two counterparts within a netting set at bank level. 

\section*{Acknowledgments}
We would like to thank Aldo Nassigh, Andrea Pallavicini and Wolfgang Runggaldier for some nice discussions on the subject. 
The usual disclaimers apply.


\section*{Appendix A}

{\it Proof of Lemma 3}. 
Function $f(\xi)$ is obtained from direct substitution of swaption payoff components (\ref{eq:B dynamics}) 
in Receiver payoff (\ref{eq:payoff}).
$f(\xi)$ is a sum of exponentials $\exp (\lambda_i \, \xi)$ multiplied by some coefficients $\omega_i$, where both $\lambda_i, \omega_i \in \Re$.
Function $f(\xi)$ is composed by different parts:
positive addends with negative exponentials (terms $a$ and $b$) and
a negative term with a positive exponential (first addend in $c$ for $\iota = \alpha'$), which becomes a negative constant for $\gamma = 0$. 
The remaining coefficients in ($c$) are always negative and they can be divided into three parts;
one with negative exponentials ($\nu_{\alpha \iota} > 0$), 
another one with positive exponentials ($\nu_{\alpha \iota} < 0$) and a third part
constant when at least one $\nu_{\alpha \iota}$ is equal to $0$.

Let us study $f (\xi)$ as a function of $\xi \in \Re$. It is a very regular function (${\cal C}^\infty$), a finite sum of exponentials and constants. 
We divide the addends of function $f$ in two parts. In the first one $f_+ (\xi)$ we consider the sum of all positive addends (i.e. terms a and b) and 
in the second one $f_- (\xi)$ the sum of all negative addends (i.e. term c) in absolute value, i.e.
\[
f (\xi) =: f_+ (\xi) - f_- (\xi) 
\]
where both $f_+ (\xi)$ and $f_- (\xi)$ are positive functions of their argument: $f_+ (\xi)$ is the sum of negative exponentials while
$f_- (\xi)$ can be the sum of both positive, negative exponentials and a constant (only for a finite set of values for $\gamma$, for the values of $\gamma$ equal to one of the $\{ {\tilde \gamma}_\iota \}_{\iota=\alpha' +1, \ldots, \omega'} $ ). 

\bigskip

First, let us observe that a positive addend is leading  for small $\xi$.
This fact is a consequence of the following inequalities that hold $\forall \iota = \alpha' + 1, \ldots, \omega'$
\be
 v_{\alpha' \iota-1} < v_{\alpha' \iota} \; , \quad \nu_{\alpha' \iota} \le (1 - \gamma) v_{\alpha' \iota} {\rm \; where \; the \; equality \; holds \; only \; for \;} \gamma = 0 \; ,
\label{eq:inequalities}
\en 
immediate consequences of volatility definitions (\ref{eq:model}).
 For all values of $\gamma$ the leading term of $f(\xi)$ for small $\xi$ is 
\[
c_{\omega} \, B_{\alpha \omega} (t_0) \, e^{-  (1- \gamma)  \, v_{\alpha \omega} \, \xi \; + \; \cdots }
\] 
because,  due to inequalities (\ref{eq:inequalities}), $- (1- \gamma)  \, v_{\alpha \omega}$ is the lowest exponent coefficient 
that multiplies $\xi$ among the exponentials  in $f (\xi)$;
i.e. there exists always a $\hat{\xi}$ s.t. $\forall \, \xi < \hat{\xi} \;\; f_+ (\xi) > f_- (\xi)$.


Then, let us define $\tilde{\gamma} := \max_\iota \tilde{\gamma}_\iota $ 
and let us  distinguish three cases depending on  $\gamma$ value:

\begin{enumerate}
\item When $ \tilde{\gamma} \le \gamma  \le 1$, $f_- (\xi)$, due to {\it Remark 2}, 
is a positive linear combination of positive exponentials (and a positive constant when $\gamma = \tilde{\gamma}$). 
Also this case admits one unique intersection with $f_+ (\xi)$, which is a sum of negative exponentials for $\gamma <1$, as mentioned above, while is a constant for $\gamma = 1$.

\item When $0 < \gamma <  \tilde{\gamma}$,  $f_- (\xi)$ is a u-shaped positive function since it is  a positive linear combination of positive and negative exponentials 
(and a constant for some values of $\gamma$).
Moreover $f_+ (\xi)$ and $f_- (\xi)$ present one unique intersection, because $f_+(\xi)$ goes to $+\infty$ for $\xi \to -\infty$
faster than $f_- (\xi)$ 
and to $0$ for $\xi \to +\infty$. 

\item The case with $\gamma = 0$ should be treated separately. In this case
\[
\begin{array}{rl}
f (\xi)  = 
& \displaystyle  \sum^{\omega}_{j=\alpha + 1} c_j B_{\alpha j} (t_0) \, e^{-  v_{\alpha j} \, \xi   -  v_{\alpha  j}^2 \, \zeta^2/2 } 
\\[1ex]
& \displaystyle  -  \beta_{\alpha'} (t_0)  - \sum^{\omega' -1}_{\iota=\alpha' +1} \left(  \beta_{\iota} (t_0) - 1 \right) \, B_{\alpha' \iota} (t_0) \, e^{- v_{\alpha'  \iota} \, \xi   -  v_{\alpha'  \iota}^2 \, \zeta^2/2}  
\end{array}
\]
all addends are negative  exponentials and constants, and then the limit for $\xi \to +\infty$ 
is equal to $-  \beta_{\alpha'} (t_0) $. 
Moreover, due to inequalities (\ref{eq:inequalities}), 
$-v_{\alpha' \, \alpha' + 1}$ (always lower than zero) is the largest exponent coefficient 
that multiplies $\xi$ among the exponentials in $f (\xi)$, 
the leading term for large $\xi$ is 
\[
- \left(  \beta_{\alpha' +1} (t_0) - 1 \right) \, B_{\alpha' \, \alpha' +1} (t_0) \, e^{- v_{\alpha'  \alpha' +1} \, \xi   -  v_{\alpha'  \alpha' +1}^2 \, \zeta^2/2}
\]
hence $f(\xi)$ tends to $-  \beta_{\alpha'} (t_0) < 0 $ from below for $\xi \to \infty$. 
With similar arguments applied to the first derivative of $f (\xi)$,  
one can show that the function has one minimum.
Summarizing, for $\gamma = 0$
the function $f  (\xi)$ is a decreasing function up to its minimum $\xi_{min}$ 
(reaching a value lower than $-  \beta_{\alpha'} (t_0) < 0$) and then it gradually goes to $-  \beta_{\alpha'} (t_0)$ from 
below for $\xi>\xi_{min}$.
Also in this case the function $f(\xi)$ presents a unique intersection with zero.
\end{enumerate}

We have then proven that, for all parameters choices,
there exists a unique value $\xi^*$ s.t $f (\xi^*) = 0$. 
The proof is complete once we observe that, for $\xi < \xi^*$, the function $f  (\xi)$ is larger than zero 
in the three cases described above $\;  \clubsuit$

\bigskip

{\it Proof of Proposition 1}. 
Due to {\bf Lemma \ref{le:unicity}}, swaption receiver is equivalent to
\[
\begin{split}
\mathfrak{R}_{\alpha \omega}(&t_0)/B (t_0, t_\alpha) 
=   \mathbb{E} \left\{ f(\xi)\right\}^+ = \mathbb{E} \left\{  f(\xi)  | \, \mathbbm{1}_{\xi \le \xi^*} \right\}  \\[1ex]
& =   \displaystyle  \sum^{\omega}_{j=\alpha + 1} c_j \, \mathbb{E} \left\{
 \left[ B_{\alpha j} (t_0) \, e^{-  \varsigma_{\alpha j} \, \xi   -  \varsigma_{\alpha  j}^2 \, \zeta^2/2 }  \right] \, \mathbbm{1}_{\xi \le \xi^*} \right\} 
+ \sum^{\omega' -1}_{\iota=\alpha' +1} \mathbb{E} \left\{
 \left[ B_{\alpha \iota} (t_0) \, e^{- \varsigma_{\alpha  \iota} \, \xi   -  \varsigma_{\alpha  \iota}^2 \, \zeta^2/2}  \right] \, \mathbbm{1}_{\xi \le \xi^*} \right\} \\[1ex] 
& \displaystyle - \sum^{\omega' -1}_{\iota=\alpha'} \mathbb{E} \left\{
 \left[ \beta_{\iota} (t_0) \, B_{\alpha \iota} (t_0) \, e^{- \nu_{\alpha  \iota} \, \xi   -  \nu_{\alpha  \iota}^2 \, \zeta^2/2} \right] \, \mathbbm{1}_{\xi \le \xi^*} \right\}  
\end{split}
\]
and then, after straightforward computations, one proves the proposition $\;  \clubsuit$

\section*{Appendix B} 

In this appendix we report the Normal-Black formula for a receiver swaption:
\[
\mathfrak{R}^\textsc{mkt}_{\alpha \omega}(t_0) = B(t_0, t_{\alpha}) \; 
 BPV_{\alpha \omega}(t_0) \; 
 \left\{ [ K -  S_{\alpha \omega}(t_0)] \; 
 N \left( - d \right) +  \sigma_{\alpha \omega}  \; \sqrt{t_\alpha -t_0} \; \phi \left( d \right)
\right\}
\]
where $N(\bullet)$ is the standard normal CDF, $\phi(\bullet)$ the standard normal density function and
$\sigma_{\alpha \omega}$ the corresponding implied normal volatility 
\[
d := \frac{ S_{\alpha \omega}(t_0) -K}{\sigma_{\alpha \omega} \; \sqrt{t_\alpha -t_0}} \; \; .
\]

The ATM formula simplifies to
\[
\mathfrak{R}^\textsc{mkt}_{\alpha \omega}(t_0) = B(t_0, t_{\alpha}) \; 
 BPV_{\alpha \omega}(t_0) \;  \sigma_{\alpha \omega}  \; \sqrt{\frac{t_\alpha -t_0}{2 \pi}} \;\; .
\]

\bibliography{illiquidity}
\bibliographystyle{tandfx}

\section*{Notation and shorthands}

\begin{center}
\begin{tabular}{|l|l|} \hline
{\bf Symbol} & {\bf Description} \\
\hline
$ a, \sigma, \gamma$ & Multicurve Hull and White (\ref{eq:model})  parameters;  $a, \sigma \in \Re^+$ and $\gamma \in [0,1]$  \\[1mm]
$B(t,T)$ & {\it discount} curve, zero-coupon bond in $t$ with maturity $T$\\[1mm]
$B(t;T, T + \Delta)$ & forward {\it discount} in $t$ between $T$ and $T + \Delta$, $t \le T < T + \Delta$ \\[1mm]
$\hat{B}(t;T, T + \Delta)$ & forward {\it pseudo-discount} in $t$ between $T$ and $T + \Delta$, $t \le T < T + \Delta$ \\[1mm]
$\beta(t;T, T + \Delta)$ & forward {\it spread} in $t$ between $T$ and $T + \Delta$, $t \le T < T + \Delta$ \\[1mm]
$\beta(t,T)$ &  {\it spread} curve  in $t$ with maturity $T$\\[1mm]
$\delta(t_j, t_{j+1})$ & year-fraction between two payment dates in swap's fixed leg \\[1mm]
$\delta(t'_\iota, t'_{\iota+1})$ & year-fraction between two payment dates in swap's floating leg \\[1mm]
$\Delta$ & the lag that characterizes the {\it pseudo-discounts}, e.g. $6$-months for Eur$6$m \\[1mm]
$K$ & strike rate \\[1mm]
$N(\bullet)$ & the standard normal CDF \\[1mm]
$\rho$ & correlation matrix in $\Re^{d \times d}$ s.t. $dW_{i, t} \; dW_{j, t} = \rho_{i\, j} \, dt$ \\[1mm]
$\sigma(t,T)$ & HJM {\it discount} volatility in $\Re^d$ between $t$ and $T$  \\[1mm]
$\eta(t,T)$ & HJM {\it spread} volatility in $\Re^d$ between $t$ and $T$ \\[1mm]
${\cal{R}}_{\alpha \omega}(t_\alpha)$ & receiver swaption payoff at expiry \\[1mm] 
$\mathfrak{R}_{\alpha \omega} (t_0)$ & receiver swaption price at value date \\[1mm]
$t_0$ & value date \\[1mm]
$t_\alpha$ & swaption expiry date \\[1mm]
$t_\omega$ & underlying swap maturity date \\[1mm]
${\bf t} := \{ t_j \}_j$ & underlying swap fixed leg payment dates, $j = \alpha +1, \ldots, \omega$ \\[1mm]
${\bf t}' := \{ t'_\iota \}_\iota$ & underlying swap floating leg payment dates, $\iota = \alpha' +1, \ldots, \omega'$ \\[1mm]
${\un W}_{\, t}$ & vector of correlated Brownian motions in $\Re^d$ s.t. $dW_{i,t} \; dW_{j,t} = \rho_{i\, j} \, dt$ \\[1mm]
$x \cdot y$ & canonical scalar product in $\Re^d$ \\[1mm]
$x^2$ & scalar product $x \cdot \rho x$ with $x \in \Re^d$ and $ \rho$ correlation matrix\\[1mm]
$\xi$ & Gaussian r.v. defined in (\ref{eq:xi}) with zero mean and variance $\zeta^2$ \\[1mm]
$\xi^*$ & the unique solution of $f(\xi) = 0$; $f(\xi)$ defined in (\ref{eq:Jam}) \\
$\zeta$ & standard deviation of the Gaussian r.v. $\xi$ \\
\hline
\end{tabular}
\end{center}

\newpage

{\bf Shorthands}

\[
\begin{array}{lcl}
B_{\alpha \, j} (t) & : &  B (t; t_{\alpha}, t_{j}) \\[1mm]
B_{\alpha' \, \iota} (t) & : &  B (t; t'_{\alpha'}, t'_{\iota}) \\[1mm]
\beta_{\iota} (t) & : &  \beta (t; t'_\iota, t'_{\iota+1}) \\[1mm]
\delta'_\iota & : & \delta (t'_\iota, t'_{\iota+1}) \\[1mm]
\delta_j & : & \delta (t_j, t_{j+1}) \\[1mm]
c_j & : & \delta_j \; K \;\;\;  {\rm for} \;\;\; j  = \alpha +1, \ldots, \omega - 1 \;\;\; {\rm and} \;\;\; 1 +   \delta_\omega \; K \;\;\; {\rm for} \;\;\; j = \omega  \\[1mm]
v_{\alpha' \, \iota} & : &  v(t_\alpha, t'_\iota)   \\[1mm]
\varsigma_{\alpha' \, \iota} & : &   (1- \gamma) \, v_{\alpha' \, \iota}   \\[1mm]
\nu_{\alpha' \, \iota} & :& \varsigma_{\alpha' \, \iota} - \left( \eta(t_\alpha, t'_{\iota+1}) - \eta(t_\alpha, t'_{\iota}) \right)  \\[1mm]
{\rm IR} & : & {\rm Interest \; Rate}  \\[1mm]
{\rm MHW} & : & {\rm Multicurve \; Hull \; White \; model} \; (\ref{eq:model})   \\[1mm]
{\rm r.v.} & : & {\rm random \; variable}   \\[1mm]
{\rm s.t.} & : & {\rm such \; that}   \\[1mm]
{\rm w.r.t.} & : & {\rm with \; respect \;  to} 
\end{array}
\; \; .
\]

\end{document}